\documentclass[%
 aip,
 amsmath,amssymb,
 reprint,%
]{revtex4-1}

\usepackage{graphicx}
\usepackage{dcolumn}
\usepackage{bm}
\usepackage[utf8]{inputenc}
\usepackage[T1]{fontenc}
\usepackage{mathptmx}
\usepackage{etoolbox}
\usepackage{color}

\usepackage{xr}
\makeatletter
\newcommand*{\addFileDependency}[1]{
  \typeout{(#1)}
  \@addtofilelist{#1}
  \IfFileExists{#1}{}{\typeout{No file #1.}}
}
\makeatother

\newcommand*{\myexternaldocument}[1]{
    \externaldocument{#1}
    \addFileDependency{#1.tex}
    \addFileDependency{#1.aux}
}

\myexternaldocument{SUPP_RESUB}

\makeatletter
\def\@email#1#2{%
 \endgroup
 \patchcmd{\titleblock@produce}
  {\frontmatter@RRAPformat}
  {\frontmatter@RRAPformat{\produce@RRAP{*#1\href{mailto:#2}{#2}}}\frontmatter@RRAPformat}
  {}{}
}%
\makeatother

\usepackage[normalem]{ulem}
\usepackage{xcolor}

\newcommand{\ADDTXT}[1]{#1}
\newcommand{\REMOVETXT}[1]{\sout{}}

\begin{document}

\title{Microwave calibration of qubit drive line components at millikelvin temperatures}

\author{Slawomir~Simbierowicz*}
\email{slawomir.simbierowicz@bluefors.com}
\author{Volodymyr~Y.~Monarkha}
\affiliation{Bluefors Oy, Arinatie 10, 00370 Helsinki, Finland}

\author{Suren Singh}
\author{Nizar~Messaoudi}
\author{Philip~Krantz}
\affiliation{Keysight Technologies, 1400 Fountaingrove Pkwy, Santa Rosa, CA 95403, USA}

\author{Russell~E.~Lake}
\affiliation{Bluefors Oy, Arinatie 10, 00370 Helsinki, Finland}



\date{\today}
\begin{abstract}
Systematic errors in qubit state preparation arise due to non-idealities in the qubit control lines such as impedance mismatch. Using a data-based methodology of short-open-load calibration at a temperature of 30 mK, we report calibrated 1-port scattering parameter data of individual qubit drive line components. At 5~GHz, cryogenic return losses of a 20-dB-attenuator, 10-dB-attenuator, a 230-mm-long 0.86-mm silver-plated cupronickel coaxial cable, and a 230-mm-long 0.86-mm NbTi coaxial cable were found to be 35$^{+3}_{-2}$~dB, 33$^{+3}_{-2}$~dB, 34$^{+3}_{-2}$~dB, and 29$^{+2}_{-1}$~dB respectively. For the same frequency, we also extract cryogenic insertion losses of 0.99$^{+0.04}_{-0.04}$~dB and 0.02$^{+0.04}_{-0.04}$~dB for the coaxial cables. We interpret the results using a master equation simulation of all XY gates performed on a single qubit. For example, we simulate a sequence of two 5~ns gate pulses (X \& Y) through a 2-element Fabry-Pérot cavity with \REMOVETXT{400-mm} \ADDTXT{276-mm} path length directly preceding the qubit, and establish that the return loss of its reflective elements must be \REMOVETXT{>9.42~dB (> 14.3~dB)}\ADDTXT{>9.7~dB (> 14.7~dB)} to obtain 99.9~\% (99.99~\%) gate fidelity. 
\end{abstract}

\maketitle 


Microwave control in a cryogenic environment is essential to quantum information processing with solid-state qubits \cite{krantz-apr-2019, bardin-ieee-2021}. The pursuit of scaling quantum systems to the level of $10^{6}$ physical qubits has motivated studies of a wide range of new microwave technologies for qubit control, qubit readout, and auxiliary systems. Consequently, rapid progress in the field of engineered quantum systems has exposed a general lack of microwave measurement and detection methods that are native to the millikelvin environment that can address its specific measurement constraints or be traced directly to existing metrological rf standards. New measurement techniques must overcome challenges such as the requirement of low operational power dissipation within the cryogenic environment, the ability to correct for the offset from the room temperature test equipment by meter-long scale interconnects, temperature-dependent impedance changes, and the practical challenge of the lack of commercially available cryogenically-compatible rf components.

In a typical qubit operation scheme, strong microwave control and readout pulses are applied at room temperature and subsequently undergo a reduction in amplitude as they are delivered to the quantum device through lossy coaxial cables and a series of thermalized cryogenic attenuators. The attenuators at each temperature stage of the dilution refrigerator measurement system are selected in order to reduce both the pulse amplitude and accompanying blackbody noise that propagates through the cable\cite{yan-prl-2018,Krinner2019}. Therefore the cryogenic attenuators and coaxial cables strongly influence the quality of the qubit control and readout pulses. Signal distortions and corresponding qubit errors due to the influence of the measurement lines have been studied using a qubit as an \textit{in\ADDTXT{-}situ} low-temperature sensor \cite{gustavsson-prl-2013, jerger-prl-2019,rol-apl-2020,chaves-apl-2021}. However, direct measurement of microwave scattering parameters (S-parameters) of individual microwave components placed at the base temperature of a dilution refrigerator measurement system remains a technical challenge due to the lack of stable and repeatable calibration standards and methodologies for accurate error extraction. The problem has been addressed in specific cases using thru-reflect-line (TRL) \cite{yeh-rsi-2013,ranzani-rsi-2013}, and 1-port short-open-load (SOL) \cite{wang-qst-2021} methods. Accurate determination of S-parameters in a cryogenic environment has wide applicability to the development of active and passive cryo-electronic devices with immediate impact on filtering \cite{Slichter-apl-2009}, interconnects\cite{solovyov-ieee-2021}, multiplexers \cite{Chapman-apl-2016}, non-reciprocal devices \cite{abdo-natcomm-2019}, single-flux-quantum-based technologies \cite{McDermott-prapp-2014,mcdermott-qst-2017}, cryo-CMOS technologies \cite{patra-ieee-2018}, quantum hardware packaging \cite{benjanin-prapp-2016,bronn-qst-2018,Huang-PRXQuantum-2021}, microwave impedance and waveform metrology \cite{Boaventura-ieee-2021}, and parametric amplifier development \cite{Macklin-Science-2015,Simoen-JAP-2015,Planat-PhysRevApplied-2019,Esposito-apl-2021,simbierowicz-rsi-2021}, among others. Beyond quantum computing, scientific applications requiring accurate S-parameter characterization at low temperature include astrophysical observations \cite{Gawande-ieee-2021} and high-frequency electronic transport experiments \cite{Pallecchi-apl-2011,Yoon-natnano-2014,Mahoney-natcomm-2017,lake-aem-2017}.

In this letter, we report direct measurements of the 1-port S-parameter of qubit drive line components over a wide frequency band at 30~mK using the 1-port SOL method. Each of the SOL calibration standards that were used at millikelvin temperature were pre-characterized at room temperature using an electronic calibration kit (ECal) \cite{jargon-ieee-2012}. The pre-characterized standards were subsequently used to define a cryogenic data-based SOL calibration at the base temperature of a dilution refrigerator measurement system \cite{Bianco-ieee-1978,wang-qst-2021}. \ADDTXT{The motivation for our data-based calibration procedure is described in more detail in the supplementary material (Sec.~\ref{sec:motivation}). In particular, this work follows the methods of Ref.~\onlinecite{wang-qst-2021}, where measurement results were corroborated by good agreement between calibrated data of cryogenic resonators to known circuit models.} By gating the data in the time-domain and analyzing reflection amplitudes, we can recover the insertion loss of components with sufficient electrical length such as coaxial cables. Time-domain gating refers to the process of selecting a region of interest in the time-domain, filtering responses outside of the region of interest, and displaying the result in the frequency domain \cite{dunsmore-mms-2008}.  We interpret the measurement results by calculating how return loss leads to problems such as qubit state preparation infidelity. The letter is organized as follows. We first present measurements of cryogenic attenuators and coaxial cables to demonstrate the typical values within a dilution refrigerator measurement system. We then discuss the results in terms of state preparation infidelity due to scattering of qubit control pulses at discrete temperature stages within the qubit drive line.

We measured components that are representative of a qubit drive line configuration in recent superconducting qubit experiments \cite{Krinner2019}. These include commercial cryogenic attenuators \ADDTXT{(specified for operation between 4~K and 398~K)} with 10-dB and 20-dB constant attenuation across the specified frequency band of operation (dc -- 18~GHz), and coaxial cables. \ADDTXT{The attenuators use an on-chip Ni–Cr thin film voltage divider resistor network \cite{simbierowicz-rsi-2021} and have SMA male and female interfaces.} Attenuation is a scalar quantity that is straightforward to obtain in a cryogenic measurement system by comparing the transmission of an attenuator-under-test with a through-reference cable using a coaxial switch. For example, it was reported in Ref.~\onlinecite{simbierowicz-rsi-2021} that a 10-dB cryogenic attenuator deviates by less than $< 0.02$~\% from its room temperature value. In contrast, determining the intrinsic reflection of a component requires accurate assignment of the reference plane in the cryogenic environment. We also studied two different types of cryogenic semi-rigid coaxial cables. The first cable was comprised of NbTi metal for both the inner conductor and outer shield. The second cable had silver-plated cupronickel center conductor with a cupronickel shield. Both coaxial cables have a nominal \REMOVETXT{design} \ADDTXT{characteristic} impedance of $50~\Omega$ with PTFE dielectric and inner conductor diameter of $d = 0.203$~mm, dielectric outer diameter of $D = 0.66$~mm, and soldered SMA connectors.

\begin{figure}[t]
\includegraphics[width=\linewidth]{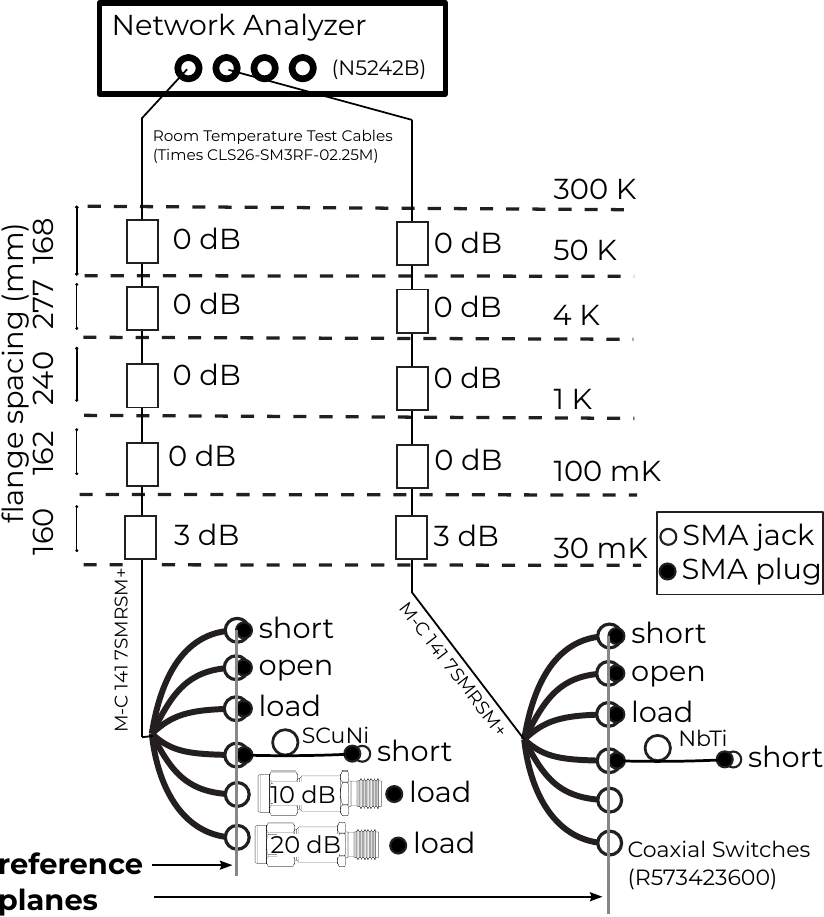}
\caption{Experimental schematic of 1-port SOL S-parameter measurements in a dilution refrigerator. Each coaxial switch enables one set of calibration standards and up to three DUTs.}
\label{fig:setup}%
\end{figure}

Figure~\ref{fig:setup} displays the schematic of the S-parameter measurement setup. The base temperature stage is equipped with two sets of coaxial switches that are used to toggle between calibration standards and each device under test (DUT). Each coaxial switch is connected to its own measurement line that is used to send and receive microwave power during \textit{in\ADDTXT{-}situ}
S-parameter characterization. Our measurement scheme enables measurements across a wide frequency band of 10~MHz -- 18~GHz by removing the added attenuation, directional couplers, circulators, and amplifiers that were present in previous schemes \cite{ranzani-rsi-2013,wang-qst-2021} with the exception of 3~dB of fixed attenuation on the mixing chamber stage. Signals were sourced and received directly to the test ports of the network analyzer (NA) using a source powers around 1~mW, and the internal electronics of the NA were used to process the data. The SOL calibration standards are installed into the system on the coaxial switches shown in Fig.~\ref{fig:setup}. We can toggle the coaxial switches to measure each standard and establish an \textit{in\ADDTXT{-}situ} reference plane in the dilution refrigerator. During the measurements we terminate the attenuators with cryogenically compatible loads that are nominally identical to the load standards, and the coaxial cables are terminated with shorts as shown in Fig.~\ref{fig:setup}. In the measurement results below, we use the convention that return loss is the negative of the reflection coefficient \REMOVETXT{expressed in decibels}, i.e., $-|\textrm{S}_{11,\textit{dB}}|$ \cite{bird-ieee-2009}.

\begin{figure*}[htb]
\includegraphics[]{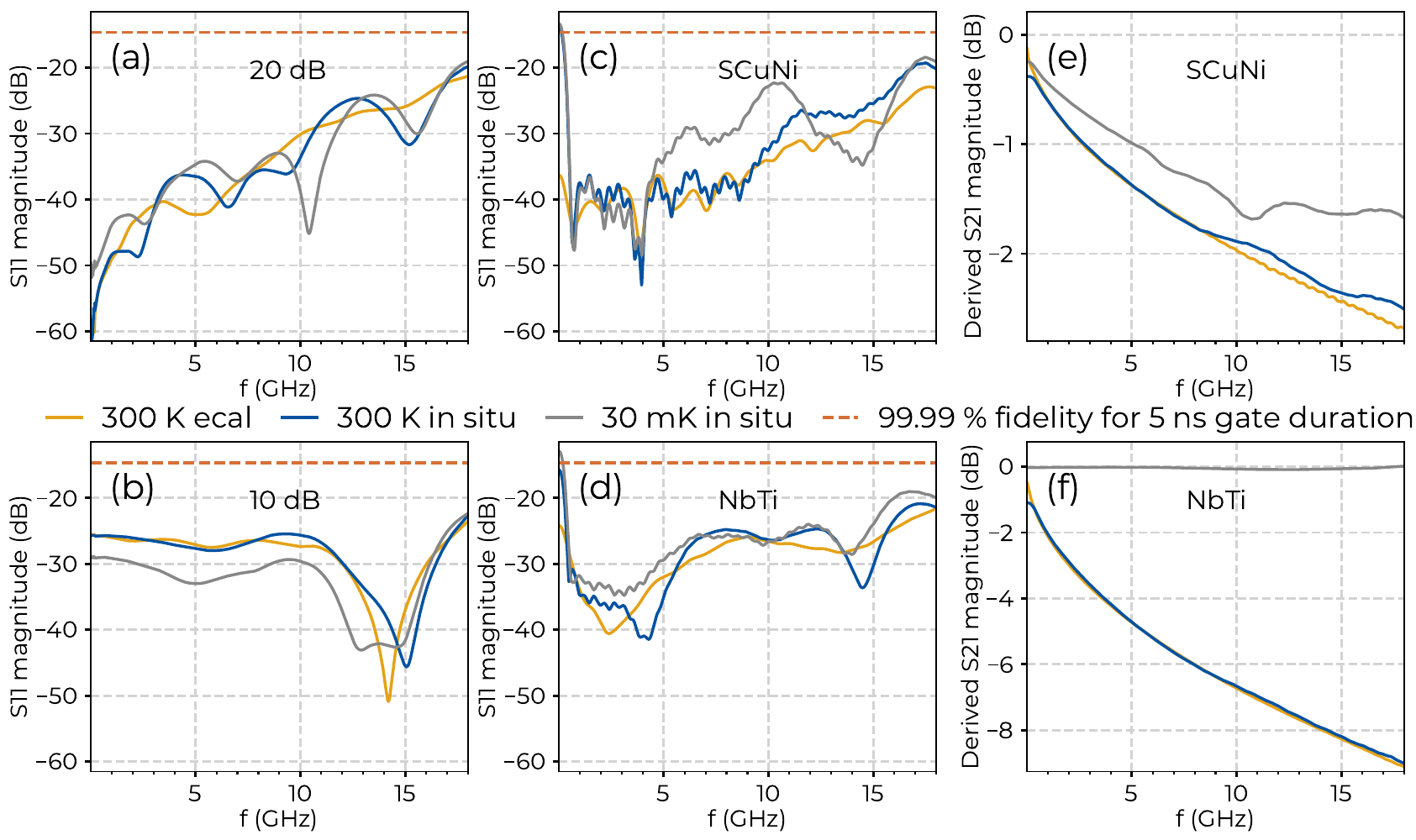}
\caption{Calibrated S-parameters \ADDTXT{of SMA-connectorized cryogenic attenuators and coaxial cables} at 300~K and 30~mK. \ADDTXT{The attenuators were always load-terminated.} All panels: Yellow traces represent 300~K measurements on the bench after a\ADDTXT{n ECal. These measurements were two-port for cable DUTs and one-port for the attenuators.}\REMOVETXT{1-port (2-port) Ecal with the VNA test cable(s) directly connected to the attenuators (coaxial cables) under test.} The blue and gray traces are SOL-calibrated data measured at 300~K and 30~mK with the DUT\ADDTXT{s} and calibration standards, in both cases, directly connected to microwave switches on the mixing chamber of the dilution refrigerator. The orange dashed lines indicate the return loss threshold for a state fidelity of 99.99~\% after 5~ns X and Y gates. \textbf{(a),(b)} Magnitude of the 1-port S-parameter for cryogenic attenuators with nominal insertion losses of 20~dB and 10~dB. \textbf{(c),(d)} \ADDTXT{Input}\REMOVETXT{Source} match of 230-mm\ADDTXT{-}long SCuNi-086 and NbTi-086 coaxial cables. \textbf{(e),(f)} Transmission magnitude of the same cables \ADDTXT{measured or} calculated from \textit{in\ADDTXT{-}situ} reflection data (see\ADDTXT{:} \REMOVETXT{the} supplementary material).}
\label{fig:sparams}
\end{figure*}

Measurement results for the frequency-dependent magnitude of the 1-port S-parameters are shown in Fig.~\ref{fig:sparams}(a)--(d). Furthermore, transmission of the cables can be derived from the reflection data using a time-gating technique to analyze the amplitude of reflections in Fig.~\ref{fig:sparams}(e),(f) (see supplementary material Sec.~\ref{TDA}). Pre-characterization of the calibration standards is performed at the ends of the test cables at room temperature (300~K) utilizing the ECal calibration. The initial measurements of each DUT are also performed at room temperature and plotted as yellow traces in Fig.~\ref{fig:sparams}. For Fig.~\ref{fig:sparams}\ADDTXT{(c)-(f)}\REMOVETXT{(e),(f)}, data plotted as yellow traces was acquired using a full 2-port S-parameter measurement at the end of the room temperature test cables. To perform the \textit{in\ADDTXT{-}situ} measurements in the dilution refrigerator, we then install the pre-characterized SOL standards into the coaxial switches as shown in Fig.~\ref{fig:setup} and upload the pre-characterization data onto the NA to define a data-based calibration kit \cite{wang-qst-2021}. We perform the \textit{in\ADDTXT{-}situ} measurements at 300~K by toggling the coaxial switches to measure each calibration standard and the DUTs (plotted as blue traces in Fig.~\ref{fig:sparams}). Finally, after measurements are performed at room temperature, the entire system is cooled down and the calibration procedure is repeated when the standards and DUTs are at 30~mK (gray traces in Fig.~\ref{fig:sparams}). 

The data plotted in Fig.~\ref{fig:sparams} has been post-processed by applying gates in the time-domain to remove any post-calibration drift of reflections that originate from higher cryostat stages and room temperature cables (see supplementary material). The geometry and materials of the coaxial cables define the characteristic impedance \REMOVETXT{match}\ADDTXT{at the center of the cable}, but the solder connection between the cable and connector is the dominant source or reflection \cite{debry-rsi-2018}. \REMOVETXT{Therefore, the data of Fig.~\ref{fig:sparams}(c),(d) was also post-processed to include only the reflection from the connector closest to the reference plane.}  \ADDTXT{Since coaxial cables have two connectors, the S-parameter data can be processed in two distinct ways: (1) applying the bandpass gate at the connector closest to the reference plane allows obtaining the return loss of Fig.~\ref{fig:sparams}(c),(d) which emphasizes the reflections at the connector, or (2) applying it to the connector with the shorting cap at the end of the cable, which isolates a signal proportional to the cable loss of Fig.~\ref{fig:sparams}(e),(f). A consequence of the gating procedure is that low frequency information is lost below the cut-off set by the inverse of the gate span 1/t$_\mathrm{gate}$ which is 200~MHz for the attenuators and $\sim$~300~MHz for the cables. Following Ref.~\onlinecite{archambeault_time_2006}, we use the original data below cut-off \cite{archambeault_time_2006} for attenuators but due to the multiple reflections within the cables, we only plot the gated data in their case.} 

Error analysis of the data presented in Fig.~\ref{fig:sparams} is described \ADDTXT{in detail} in the supplementary material. \ADDTXT{In summary, we have investigated and quantified}\REMOVETXT{The} error sources \ADDTXT{such as (i)}\REMOVETXT{include} the reflection uncertainty from a root sum squared (RSS) analysis of error sources inherent to the ECal \ADDTXT{obtained with a calculator from Ref.~\onlinecite{keysight_downloadable_nodate}}, \ADDTXT{(ii) switch variability: }\REMOVETXT{as well as} the difference in the transmission and reflection between microwave switch ports\ADDTXT{, (iii) switch repeatability: the average change in reflection magnitude for a switch port after four switching events, and (iv) differences in the reflection magnitude of the loads used to terminate the attenuators for measurements of Fig.~\ref{fig:sparams}. When calculating the total error as an RSS sum, we find that the errors (i) and (ii) are sufficient to represent the error, as (iii) and (iv) would only contribute a maximum of 2~\% and 0.04~\% to the total error and are omitted. The resulting error bars, representing 64~\% confidence intervals, are equal in linear units but look asymmetric when converted to logarithmic units with $RL(\text{dB}) = -20\log_{10}(|S_{11,\mathrm{lin}}|~\mp~\sigma_\mathrm{RSS})$. Here, $RL(\text{dB})$ is the return loss, $S_{11,\mathrm{lin}}$ the gated one-port S-parameter data in linear units, and $\sigma_\mathrm{RSS}$ the combined RSS error}.

When observing both room temperature and cryogenic data in Fig.~\ref{fig:sparams}(a), we note that the return loss ($=-|S_{11}|$) of the 20-dB-attenuator is above 40~dB at low frequencies and stays above 30~dB until around the frequency of 10~GHz and reaches its minimum value of 19~dB at 18~GHz. There is not a significant temperature dependence aside from the resonance at 10.4~GHz that only appears at 30~mK. At the frequency of 5~GHz\ADDTXT{,} the \textit{in\ADDTXT{-}situ} 300~K and 30~mK return loss values are \REMOVETXT{33}\ADDTXT{36}$^{+4}_{-3}$~dB \ADDTXT{(0.015~$\mp$~0.006)} and 35$^{+3}_{-2}$~dB \ADDTXT{(0.019~$\mp$~0.006)} respectively\ADDTXT{, where the corresponding values in linear units are in parentheses. The error bars are always dominated by the switch variability error which does not depend on the magnitude of reflection and therefore at 5~GHz the error is roughly 0.006 for all devices}. On the other hand, for the 10-dB-attenuator displayed in Fig.~\ref{fig:sparams}(b), the return loss increases at cryogenic temperatures by 3--5~dB with respect to the room temperature measurements up until a frequency of approximately 10~GHz. At higher frequencies of $\sim 15$~GHz the measured return loss values at 300~K and 30~mK converge. At the selected frequency of 5~GHz, the \textit{in\ADDTXT{-}situ} room temperature return loss value is 28$^{+1}_{-1}$~dB \ADDTXT{(0.041~$\mp$~0.006)}, and the 30~mK measurement result is 33$^{+3}_{-2}$~dB \ADDTXT{(0.022~$\mp$~0.006)}. 

As expected, the cryogenic insertion losses of Fig.~\ref{fig:sparams}(e),(f) are reduced with respect to the room temperature data. For example at a frequency of 5~GHz, the \textit{in\ADDTXT{-}situ} loss 1.38$^{+0.04}_{-0.04}$~dB of the SCuNi cable decreases to 0.99$^{+0.04}_{-0.04}$~dB when only switch and calibration errors are considered. The \textit{in\ADDTXT{-}situ} room temperature measurement (blue) agrees within $\pm$0.2~dB with the data obtained with ECal (yellow) where the dropoff at low frequencies we attribute to the time gate \cite{archambeault_time_2006}. For the NbTi coaxial cable, the loss measured \textit{in\ADDTXT{-}situ} at 5~GHz drops from 4.72$^{+0.03}_{-0.03}$~dB to 0.02$^{+0.04}_{-0.04}$~dB. The remaining errors are similar to those observed for SCuNi, however, a larger gate-induced difference of 0.4~dB is seen at the lowest measured frequencies. A tabulation of return loss measurement results for the 10-dB and 20-dB attenuators, and the SCuNi and NbTi coaxial cables is shown in Table~\ref{tab:returnloss}.

\begin{table}[b]
\caption{\label{tab:returnloss}Return loss at 30~mK of tested components at select frequencies. The error bars represent the combined error of ECal uncertainty and variability of RF switch ports (see supplementary material).\ADDTXT{(*) Only the lower bound is known.}}
\begin{ruledtabular}
\begin{tabular}{cccccccc}
 &\multicolumn{6}{c}{Return loss (dB)}\\
\textbf{DUT} &1~GHz&2~GHz&4~GHz&5~GHz&8~GHz&16~GHz\\ \hline
\\[-0.8em]
\textbf{20 dB} &44$^{+3}_{ -2}$&42$^{+5}_{ -3}$ &37$^{+3}_{ -2}$ &35$^{+3}_{ -2}$ &35$^{+4}_{ -3}$ &28$^{+4}_{ -3}$   \\ 
\\[-0.8em]
\textbf{10 dB} &29$^{+1}_{ -1}$&30$^{+1}_{ -1}$ &32$^{+2}_{ -1}$ &33$^{+3}_{ -2}$ &31$^{+3}_{ -2}$ &31$^{+7}_{ -4}$   \\ 
\\[-0.8em]
\textbf{SCuNi} &\ADDTXT{39$^{+2}_{ -1}$}\REMOVETXT{27.9$^{+0.5}_{ -0.4}$}&\ADDTXT{41$^{+4}_{ -3}$}\REMOVETXT{40$^{+4}_{ -3}$}&\ADDTXT{48$^{+\infty}_{ -6}$*}\REMOVETXT{39$^{+5}_{ -3}$}&34$^{+3}_{ -2}$ &30$^{+2}_{ -2}$ &24$^{+2}_{ -2}$   \\ 
\\[-0.8em]
\textbf{NbTi} &\ADDTXT{32$^{+1}_{ -1}$}\REMOVETXT{26.5$^{+0.4}_{ -0.4}$}&\ADDTXT{33$^{+1}_{ -1}$}\REMOVETXT{34$^{+2}_{ -1}$} &\ADDTXT{33$^{+2}_{ -2}$}\REMOVETXT{32$^{+2}_{ -1}$} &29$^{+2}_{ -1}$ &26$^{+1}_{ -1}$ &\REMOVETXT{21}\ADDTXT{20}$^{+1}_{ -1}$  
\\ 
\\[-0.8em]
\end{tabular}
\end{ruledtabular}
\end{table}

\begin{figure}[htb]
\includegraphics[width=\linewidth]{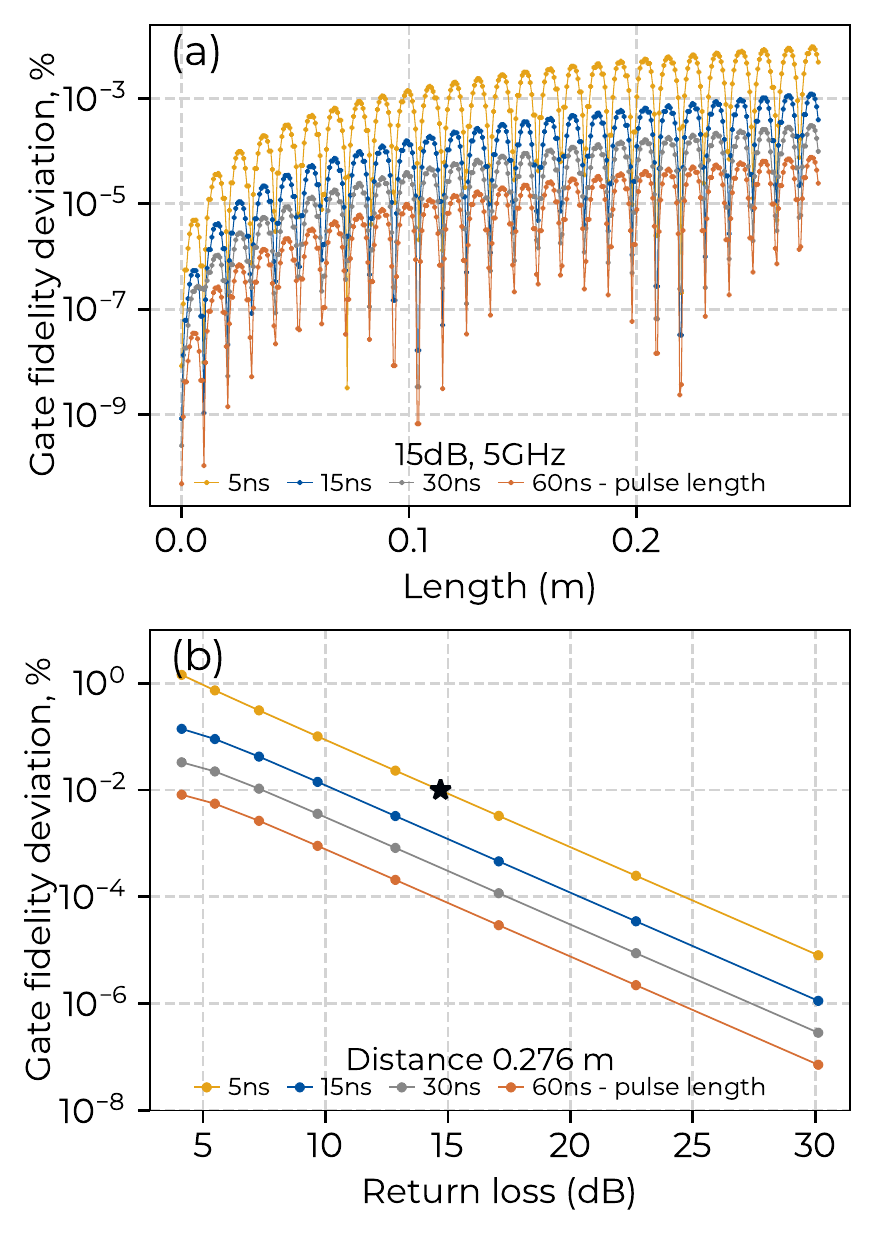}
\caption{Numerically simulated fidelity deviation for an XY gate combination due to applied gate pulse distortion depending on the length (a) of the transmission line between two unmatched elements for a fixed return loss value of 15~dB and on the return loss (b) at a fixed \REMOVETXT{length of 0.4~m} \ADDTXT{distance of 0.276~m between unmatched elements (close to a maximum in fidelity deviation)} at a frequency of 5~GHz. Parameters for the return loss threshold plotted in Fig.~\ref{fig:sparams}(a)--(d) are indicated with a black star.
\label{fig:fidelity}}%
\end{figure}

We now turn to a discussion of the influence of the return loss results of Table~\ref{tab:returnloss} on single-qubit gate fidelity. In microwave pulse driven qubit architectures single-qubit gate ﬁdelity becomes highly susceptible to any impedance mismatch in the microwave line since such imperfections lead to pulse distortions \cite{gustavsson-prl-2013}. We present a model where a qubit drive pulse travels through a transmission line with two unmatched elements \ADDTXT{ (representing cable connectors or poorly matched attenuators) } that partially reflects between these elements in a way similar to Fabry–Pérot interferometer. The resulting reflections interfere with the original pulse causing distortions in both phase and amplitude.   

The numerical simulation is performed using Qutip \--- an open-source python library for simulating the dynamics of open quantum systems \cite{qutip_2} with additional details reported in the supplementary material Sec.~\ref{supsec:sim-details}. Within the simulation we perform an ALLXY benchmarking set \cite{gao-prxq-2021}. The end result is the fidelity between the state achieved for an ideal case of infinite return loss and the one achieved with a given finite return loss. The results of the simulations are presented in Fig.~\ref{fig:fidelity}. In addition, we plot the return loss threshold of \ADDTXT{14.7~dB}\REMOVETXT{14.3~dB} in Fig.~\ref{fig:sparams}(a)--(d) that corresponds to 99.99~\% state fidelity for X and Y gates applied to a qubit in sequence with a 5-ns gate-pulse through a 276-mm cable.

As expected, the simulation results reveal that the fidelity deviation $1-F$ depends periodically on the length between the two unmatched elements with the period defined by pulse frequency and return loss. For any return loss one can find a suitable length between unmatched elements of signal frequency where the fidelity deviation induced by the effect under study is minimal. From the simulation results we observe that gate fidelity deviation becomes stronger for shorter drive pulse duration\ADDTXT{s}. For example, from Fig.~\ref{fig:fidelity} we observe that fidelity deviation for 60~ns pulses and 5~ns pulses differ by two orders of magnitude. Thus\ADDTXT{,} the effect may become relevant in experimental systems when working with short drive pulses \cite{werninghaus_leakage_2021}.

Fidelity deviations arise due to the phase change within the drive pulse due to interference of the original signal with the delayed reflections. In turn, this phase changes lead to changes of the rotation angle of the qubit state vector on the Bloch sphere during gate evolution. This \ADDTXT{phenomenon}\REMOVETXT{phenomena} can be described by a curving of the trajectory of the qubit state on the Bloch sphere. That curvature can be positive or negative depending on the phase differences between the original signal and the reflections. The effect is most prominent near the beginning and the end of the drive pulse, where multiple phase changes are present.

In conclusion, we present calibrated 1-port S-parameter measurements of 10-dB and 20-dB cryogenic attenuators, and two different varieties of cryogenic coaxial cables to verify impedance matching. Our measurement technique removes all excess attenuation and frequency-dependent components within the cryogenic environment in order to measure reflections directly using a high-dynamic range NA over a broad frequency band of 10~MHz--18~GHz with pre-characterized data-based calibration standards \textit{in\ADDTXT{-}situ}. We also present a theoretical analysis of gate fidelity deviation as a function of return loss of the attenuators, cable length, and gate-pulse duration. For short cable lengths, larger return loss values can be tolerated and mismatch-induced pulse distortions will not be a significant source of single-qubit gate errors.  Decreasing the gate-pulse duration increases the return loss requirement for maintaining high gate-fidelity. The results highlight the importance of keeping cables between mismatched elements as short as possible in order to avoid single qubit preparation errors. This point is crucial when designing larger modular multi-chip systems that are connected via coaxial cables in scaled-up quantum processors, or systems that have large shielded enclosures that necessitate longer interconnects.


\section*{Supplementary material}
See supplementary material for detailed descriptions of the calibration procedure, time domain and error analysis, and master equation simulations.

\section*{Acknowledgements}
V.Y.M. would like to thank Dr. Aidar Sultanov from Aalto University for helpful discussions regarding master equation simulations.

\section*{Authors' contributions}
S.~Simbierowicz and V.Y.M. contributed equally to this work.

\section*{Data availability}
All data that support the findings of this study are
available from the corresponding author upon reasonable request.

\bibliography{2portcal}

\end{document}


\preprint{AIP/123-QED}

\title{Supplementary Material for "Microwave calibration of qubit drive line components at millikelvin temperatures"}

\author{Slawomir~Simbierowicz*}
    \email{slawomir.simbierowicz@bluefors.com}
\author{Volodymyr~Y.~Monarkha}
\affiliation{Bluefors Oy, Arinatie 10, 00370 Helsinki, Finland}

\author{Suren~Singh}
\author{Nizar~Messaoudi}
\author{Philip~Krantz}
\affiliation{Keysight Technologies, 1400 Fountaingrove Pkwy, Santa Rosa, CA 95403, USA}

\author{Russell~E.~Lake}%
\affiliation{Bluefors Oy, Arinatie 10, 00370 Helsinki, Finland}

\date{\today}
\maketitle

\section{Experimental Methods}

\subsection{Calibration procedure}
\label{procedure}

Measurements were performed in a Bluefors XLDsl dilution refrigerator measurement system with a Keysight vector network analyzer (N5242B). The procedure can be briefly summarized in the following five steps. First, we calibrate the VNA at the end of the pair of the armored room temperature test cables (Times Microwave CLS26-SM3RF-02.25M) \ADDTXT{, with high-quality SMA bulkhead adapters,} using an electronic calibration kit (N4692A). Second, we measure the SMA male calibration standards and using these measurements we define a databased-calibration kit as in Ref.~\citenum{wang-qst-2021}. Third, we connect the same standards and DUTs to the microwave switches in the mixing chamber of the dilution refrigerator and cool down the system. Fourth, when the system has reached base temperature we use the guided routine and the defined databased calibration kit on the network analyzer to perform the calibration actuating the rf switch as necessary. We always wait that the system cools down to 30~mK after a switching event. Finally, we measure the DUTs on the other switch ports using the correction from the guided calibration process used in the previous step.

Importantly, we note that we always kept the VNA settings the same for measurements that followed calibration although slight variations to settings were allowed between measurement runs. However, the critical parameters of frequency range or attenuation settings used in the calibration of the VNA were not changed during the measurement. This was done to ensure the integrity of calibration used for measurements was not compromised. Therefore, the uncertainties we derive in Sec.~\ref{sec:Errors} that rely on the calibration were not altered during the measurement of the devices. Specifically, we always used a frequency window of 10~MHz to 26.5~GHz on the VNA with 10001~points for the standards and attenuators whereas 10597~points were used in the case of coaxial cables since the measurements were performed during multiple cooldowns.  \ADDTXT{For the measurements of the cables, The number of points was selected so that the start frequency was equal to an integer multiple of the frequency step, i.e., $f_\textrm{start} = n\Delta f = 4\cdot2.5$~MHz, to simplify certain time domain and de-embedding operations. None of those operations are described in this study. Therefore the number of points was arbitrary, but consistent throughout the measurements of the coaxial cables and included a relatively large number of points.} Additionally, the room temperature measurements of standards and attenuators were performed using a VNA probe power of 0~dBm, and the cables at -10~dBm. Changing power within this range did not actuate the step attenuator on the VNA and therefore we do not expect significant error from the power change. During the cryogenic measurements, we used a probe power of -10~dBm instead to prevent the mixing chamber from heating up during measurements.

\subsection{Time-domain analysis}
\label{TDA}
\begin{figure}[htb]
\includegraphics[width=\linewidth]{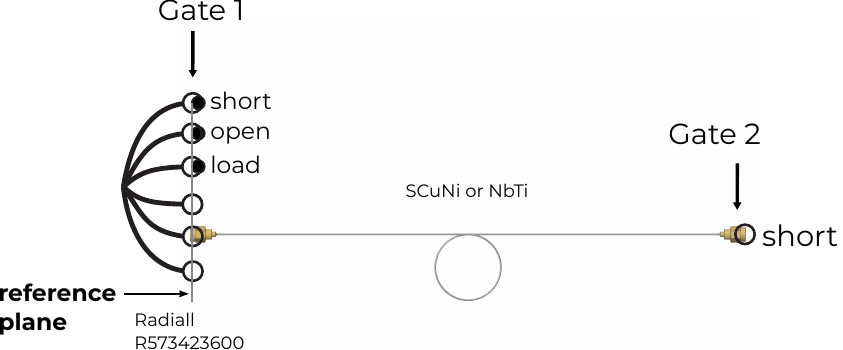}
\caption{Illustration of time gates applied to the coaxial cable (not to scale) installed on the microwave switch. Applying a \ADDTXT{3~ns}\REMOVETXT{1~ns} wide bandpass gate at "Gate~1", or equivalently at a time delay of 0~ns, gives the reflection magnitude of the cable connector on the switch. Respectively, a\REMOVETXT{n identical} \ADDTXT{3.8~ns wide} gate at "Gate~2", or 2.15~ns, gives twice the transmission magnitude of the whole cable. \label{fig:gates}}%
\end{figure}

\ADDTXT{The S-parameter data displayed in Fig.~\ref{fig:sparams} of the main text have been post-processed using band-pass gates as we will detail below. A summary of the gate parameters is included in Table~\ref{tab:gates}.} The S-parameter data for attenuators in the main text Figs.~\ref{fig:sparams}(a) and (b) have been gated with wide \ADDTXT{t$_\mathrm{gate}$~=}~5~ns bandpass gates at the reference plane simply to remove any post-calibration drift that originates from higher cryostat stages and room temperature cables. \ADDTXT{The gate imposes a frequency cut-off 1/$t_\mathrm{gate}$ = 200~MHz to the data below which original ungated data may be used \cite{archambeault_time_2006} and we have done so in Fig.~\ref{fig:sparams}(a) and (b).} \REMOVETXT{The S-parameter data for cables, however, is processed with more localized gates as illustrated in Fig.~\ref{fig:gates}. Specifically, the reflection data of Figs.~\ref{fig:sparams}(c) and (d) was obtained with a 1~ns bandpass gate applied at the reference plane (Gate~1).} 

\ADDTXT{Processing the S-parameter data for cables, to obtain Fig.~\ref{fig:sparams}(c)-(f), required additional steps.}\REMOVETXT{The transmission magnitude in Fig.~\ref{fig:sparams}(e),(f) can be derived from the calibrated reflection data in the following way.}  After frequency dependent $|S_{11}|$ data was acquired, the data was transformed to the time-domain using an inverse Fourier transform. \ADDTXT{In this way, reflection peaks in time could be identified and we could then say with certainty that the reflections take place at the ends of the cable. Illustrated in Fig.~\ref{fig:gates}, the obtained time-domain data could then be processed by applying a 3~ns bandpass gate at the reference plane (Gate 1) and, converting back to the frequency domain, yielded the reflection data of Figs.~\ref{fig:sparams}(c) and (d). Importantly, this procedure excludes the reflection at the short corresponding to 2.15~ns~=~$2Lv_{\textrm{p}}^{-1}$, where $L$~\ADDTXT{=~230~mm} is the length of the cable, and $v_{\textrm{p}}=0.7c$ is the phase velocity in the coaxial cable with Teflon dielectric in terms of the speed of light in vacuum $c$.}
\REMOVETXT{The reflection peaks in time were identified and a 1~ns bandpass gate was applied at the time corresponding to the shorting cap (Gate 2) at the end of the cable.} 

\ADDTXT{The transmission magnitude in Fig.~\ref{fig:sparams}(e),(f) can be derived from the calibrated reflection data in the following way. In contrast to what we did above, we may process the time-domain transformed cable data by applying a gate at Gate 2 isolating the reflection from the short. Incidentally, we may use a wider 3.8~ns bandpass gate as we are not that concerned about the low reflection at the reference plane. Yet, we still keep t$_\mathrm{gate}$ below 4~ns.} \REMOVETXT{This procedure isolates the reflection corresponding to $2Lv_{\textrm{p}}^{-1}$, where $L$ is the length of the cable, and $v_{\textrm{p}}=0.7c$ is the phase velocity in the coaxial cable with Teflon dielectric in terms of the speed of light in vacuum $c$. Importantly, the cable is longer in time $\sim 2$~ns than the span of the time gate so as to capture only the intended reflection.} The gating procedure is based on the following idealizations: (1) nothing reflects at the reference plane, (2) all power that reaches the short is reflected, and (3) the remaining signal arrives at the input without additional reflections taking place. This means all the loss is insertion loss from the component. Therefore, we find the relation
$|S_{21}||S_{12}| = |S_{11,\text{gated}}|$
where $S_{21}$ and $S_{12}$ are the two-port S-parameters of the cable, and $|S_{11,\text{gated}}|$ corresponds to the fraction of signal going through the component, reflecting at the short, and traveling back to the input. Here, assuming that the cable is reciprocal meaning $|S_{21}| = |S_{12}|$ we may obtain the transmission magnitude of the component $|S_{21}|$ by taking the square root on both sides. We arrive at the equation
\begin{equation}
|S_{21}| = \sqrt{|S_{11,\text{gated}}|} \text{,}
\end{equation}
which we may now use to derive the transmission magnitudes for Figs.~\ref{fig:sparams}(e),(f).

\begin{table}[b]
\caption{\label{tab:gates}\ADDTXT{Summary of gate parameters. The gate center position is with respect to the reference plane. The bandpass gates utilize Kaiser windowing with a $\beta$-value of 6.}}
\begin{ruledtabular}
\begin{tabular}{ccccc}
 &\multicolumn{4}{c}{}\\
\ADDTXT{\textbf{Fig.~\ref{fig:sparams}} panels} & \ADDTXT{Gate type} & \ADDTXT{Center position (ns)} &\ADDTXT{Span (ns)}\\ \hline
\\[-0.8em]
\ADDTXT{\textbf{a,b}} & \ADDTXT{bandpass} & \ADDTXT{0} & \ADDTXT{5} &  \\ 
\\[-0.8em]
\ADDTXT{\textbf{c,d}} & \ADDTXT{bandpass} & \ADDTXT{0} & \ADDTXT{3} &  \\ 
\\[-0.8em]
\ADDTXT{\textbf{e,f}} & \ADDTXT{bandpass} & \ADDTXT{2.15} & \ADDTXT{3.8}  &  \\ 
\\[-0.8em]
\end{tabular}
\end{ruledtabular}
\end{table}

\subsection{Error \ADDTXT{e}\REMOVETXT{E}stimates}
\label{sec:Errors}

The data-based calibration method exhibits less systematic error due to standard definitions than polynomial models \cite{Keys-app-note-2021}. Previous studies have found that the largest source of systematic error are often the subtle differences in return loss of microwave cables that branch out to the standards and DUTs from the rf switches \cite{yeh-rsi-2013, ranzani-rsi-2013}. The worst-case error occurs when the DUT path has the largest difference from the calibration paths \cite{agilent-app-note-1998,ranzani-rsi-2013}. A common method \cite{yeh-rsi-2013, ranzani-rsi-2013} is to use phase-matched cables and even screen the cables for closely-matched reflection response. Here, we have omitted the cables which eliminates the error from cable to cable variation between calibration and measurement. Instead, this error reduces to a reflection and transmission difference between the rf switch ports. 

In the main text, we assign a 64\% confidence interval for the return loss values of Table~\ref{tab:returnloss} using the method of residual sum of squares (RSS). Following the principles of Ref.~\onlinecite{agilent-app-note-1998}, we get
\begin{equation}\label{eq:err_rl}
\begin{split}
    |S_{11,m}| &= |S_{11,a}| \pm \sqrt{\sigma_{S_{11},\text{ECal}}^2+\sigma_{S_{11},\text{switch}}^2+(|S_{21,a}|^2\sigma_{S_{11},\text{load}})^2} \\
    & \ADDTXT{\text{$\approx |S_{11,a}| \pm \sqrt{\sigma_{S_{11},\text{ECal}}^2 +\sigma_{S_{11},\text{switch,var}}^2}$}} \text{,}
\end{split}
\end{equation}
where $S_{ij,m}$ and $S_{ij,a}$ refer to measured and actual S-parameters with $i$ and $j$ indicating the port parameter. The $\sigma_{S_{11},x}$ are standard errors as defined below.

The standard error in ECal $\sigma_{S_{11},\text{ECal}}$ affects the standard definitions and therefore it can be used as a first-order estimate for the uncertainty in the data. It is estimated with a freely available uncertainty calculator offered by Keysight \cite{keysight_downloadable_nodate} selecting the RSS method within the application. \ADDTXT{The output is a table of uncertainties against magnitude of the S-parameter and from these tables the errors at specific $S_{11}$ measurement values are obtained via interpolation.} The second standard error $\sigma_{S_{11},\text{switch}}$ is estimated by a room temperature two-port measurement of the six-port switch that was performed after an ECal routine. The VNA test cables were directly connected to the common port of the switch and, one at a time, the ports \ADDTXT{where}\REMOVETXT{that housed} the DUTs \ADDTXT{had been} during the SOL measurements. The switch was opened and closed back to the same switch port \ADDTXT{four}\REMOVETXT{multiple} times and the same procedure was then repeated for the other ports. In this way, we may calculate the difference in magnitudes to the initial value of $S_{11}$ for the six switch ports included in the analysis. In the same run, we also measured the standard ports on one switch allowing us to calculate the switch error \ADDTXT{$\sigma_{S_{11},\text{switch,var}}$}\REMOVETXT{$\sigma_{S_{11},\text{switch}}$} as the standard deviation of the $S_{11}$ traces. We find that the repeatability error \ADDTXT{$\sigma_{S_{11},\text{switch,rep}}$ would increase the total RSS error of Eq.~\eqref{eq:err_rl} by a maximum of 2~\% if it were included in the switch error term as $\sigma_{S_{11},\text{switch}}^2=\sigma_{S_{11},\text{switch,var}}^2+\sigma_{S_{11},\text{switch,rep}}^2$ and the ECal uncertainty was evaluated at a demanding level of $|S_{11}| = -50$~dB.} \REMOVETXT{is never more than 23~\% of the variability error except below 300~MHz it reaches a maximum value of 66\%.} Therefore, we do not factor the repeatability error in the calculations. For example at 5~GHz, the variability error found in the way we just described is $5\cdot 10^{-3}$ and the repeatability error is $4\cdot 10^{-4}$. 

\begin{figure}[htb]
\includegraphics[width=\linewidth]{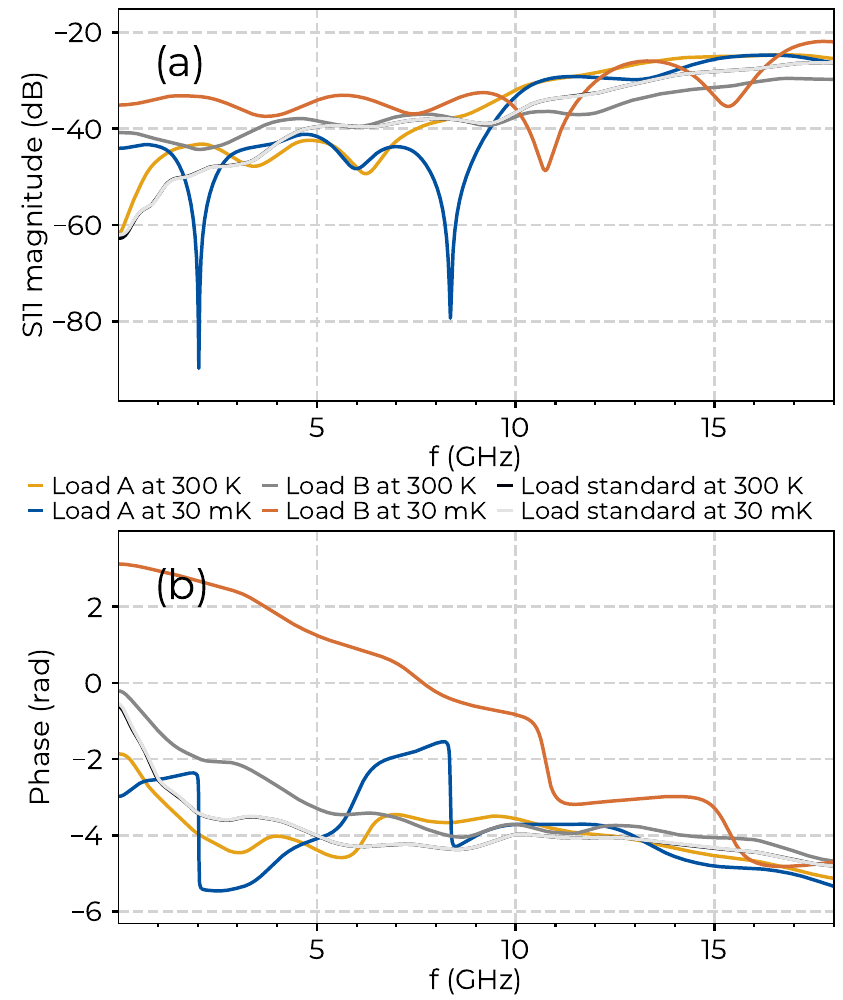}
\caption{SMA Male Loads at 300~K and 30~mK. The room temperature measurements were performed after a 1-port ECal at the end of the VNA test cable, and the 30~mK measurements \textit{in\ADDTXT{-}situ} after an SOL calibration with the displayed load standard. For the \textit{in\ADDTXT{-}situ }measurements of main text Fig.\ref{fig:sparams}, the loads A and B were used to terminate the 10-dB and 20-dB attenuators, respectively. \label{fig:loads}}%
\end{figure}

To evaluate the error term $\sigma_{S_{11},\text{load}}$ of Eq.~\eqref{eq:err_rl} we perform a supplementary measurement of the cryogenically compatible male SMA loads that are used in the main article. After a 1-port ECal at the end of the test cable at 300~K, the standard and the loads are connected to the test cable and measured one after another. We connect the standard back to the rf switch in the dilution refrigerator and instead of the DUT devices we cool down the loads. In Fig.~\ref{fig:loads}, we display the results for both 300~K and 30~mK temperatures for the standard, and the loads that were used (for measurements of Fig.\ref{fig:sparams}) to terminate the 10-dB (load A) and 20-dB-attenuators (load B). We note that the return losses are better than 20~dB through the whole measurement band from 10~MHz to 18~GHz. However, the error $\sigma_{S_{11},\text{load}}$ pertains to the differences in reflection magnitude between the loads. For the 10-dB-attenuator the prefactor $|S_{21,a}|^2$ is approximately $0.1^2$ which makes the error term even smaller than the repeatability term with \ADDTXT{only a projected 0.04~\% maximum increase to the total error.}\REMOVETXT{only a 5\% fraction of the variability term above 300~MHz and only spikes above 20\% below 80~MHz.} This error term is even smaller for the 20-dB attenuator due to its higher loss. Therefore, we do not include the load error in our calculations for error bars. In reality, the switch variability and the load variability are both systematic errors in nature and their effects could be removed from the data although we have not done so here. On another note, we have not included any errors caused by time-domain analysis.

We perform a similar analysis on the errors governing the derived insertion loss of main text Figs.~\ref{fig:sparams}(e) and (f) using
\begin{equation}
    |S_{21,m}| = |S_{21,a}|\cdot(1 \pm \sqrt{\sigma_{S_{11},\text{ECal}}^2+(2\sigma_{S_{21},\text{switch})
    })^2}) \text{,}
\end{equation}
where the switch error $\sigma_{S_{21},\text{switch}}$, contrary to Eq.~\eqref{eq:err_rl} for return loss, is instead obtained from the magnitudes of the $S_{21}$-parameters collected during the room temperature switch variability experiment. We multiply the $S_{21}$ error term by two since we are performing a two-port extraction from a one-port measurement as we discussed in Sec. \ref{TDA}. 

\subsection{\ADDTXT{Validity of data-based calibration at 30~mK}\REMOVETXT{Calibration repeatability}}\label{sec:motivation}

\ADDTXT{There are two common methods used to describe calibration standards: polynomial calibration that uses a set of closed-form polynomial equations to describe the physical properties of the standards, and data-based calibration that uses the measurement data directly \cite{wang-qst-2021}. We chose data-based calibration due to empirical observations of Ref.~\onlinecite{wang-qst-2021}, that revealed a more accurate representation of calibration standards at cryogenic temperature. Specifically, in the study associated with Ref.~\onlinecite{wang-qst-2021}, polynomial standard definitions were used in the cryogenic environment and significant deviations were observed between the model and physical standards. In contrast data-based definitions produced expected results. i.e., the S-parameters of superconducting resonator devices were well-described by the standard circuit models. Therefore, our study follows the data-based calibration approach, where data-based definitions are acquired at room temperature.}

\begin{figure}[htb]
\includegraphics[width=\linewidth]{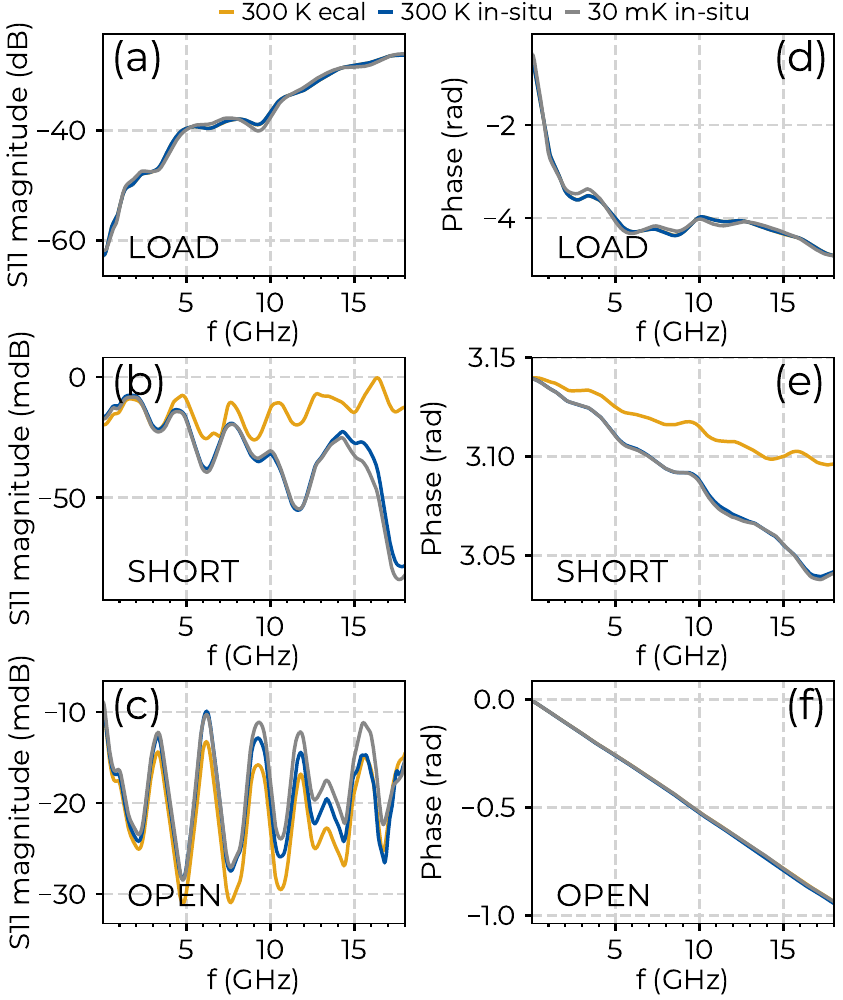}
\caption{Standards measured at 300~K (yellow) on the bench after an ECal, and \textit{in\ADDTXT{-}situ} after SOL calibration at 300~K (blue) and 30~mK (gray). \label{fig:standards}}%
\end{figure}

\ADDTXT{As an indirect study of validity, we first} \REMOVETXT{We} check the repeatability of the databased calibration method by measuring the standards after performing a calibration and display the results in Fig.~\ref{fig:standards}. The yellow traces \ADDTXT{that can be traced to the ECal} comprise the data used for defining the standards, which we call pre-characterization in the main text. We use these data as the cal kit definitions for \ADDTXT{the} subsequent \ADDTXT{\textit{in\ADDTXT{-}situ}} calibrated measurements at 300~K and 30~mK that are shown in blue and gray, respectively. We see that any deviations in measured magnitudes are in the mdB scale, and in the phase only the short deviates from the ECaled yellow trace by a maximum of 0.06~radians at 18~GHz. \ADDTXT{Furthermore, we observe distinct reflection behaviour at 30~mK between the S, O, and L standards to support the hypothesis that the load does not fail.}\REMOVETXT{Therefore, we conclude that the databased method finds the correct scattering parameters for the standards both at room temperature and cryogenic temperatures, and allows correct calibration of the DUTs.}

\ADDTXT{In addition to the repeatability study, to gain confidence in our load standard, we determine the temperature-dependent resistance change of the load by measuring a nominally identical sample in a 4-wire-resistance measurement in a separate cooldown. The measurement results revealed R$_{300\mathrm{K}}$ = 49.86~$\Omega$ and R$_{30 \mathrm{mK}}$ = 51.26~$\Omega$ for room temperature and base temperature respectively. This corresponds to a relatively small change of $\Delta R/R_{\mathrm{300K}} \sim 2.8 \%$ change where $\Delta R = R_{30 \mathrm{mK}} - R_{300 \mathrm{K}}$, and provides confidence that the reference impedance was approximately 50~$\Omega$ even at 30~mK.}

\ADDTXT{From the data of Fig.~\ref{fig:standards}(e) and (f) we may also extract the offset delay of the short and open. We observe that their absolute values are small enough (<~1~ps, and 4~ps respectively at room temperature), that even when considering the possibility of a highly exaggerated thermal contraction, the electrical length of the short and open would contribute to a negligible shift in the calibration plane, or in the loss that a small section of air gap causes. Additionally, while performing the time-domain transformations to the data for coaxial cables of Fig.~\ref{fig:sparams}(c) and (d), we observed that the hypothetical small changes to the standards with respect to room temperature values do not translate to visible changes in the locations of the reflection peaks.}

\ADDTXT{In the future, accuracy can be improved by performing detailed dimensional metrology of the mechanical sub-components of the standard and modeling of the thermal contractions. However, the contractions will only be a few percent and affect all of the standards. The thermal contraction presents a systematic error that will need to be quantified in future work. See, for example method of Ref.~\onlinecite{wong-1992}.}

\subsection{Simulation details}
\label{supsec:sim-details}

We perform a numerical simulation of the distortions introduced to the qubit drive pulse due to reflections, and the corresponding deviation of the resulting qubit state. The simulation process can be separated in two parts: (1) simulation of a sequence of distorted Gaussian shaped drive pulses\cite{ku_single_2017}, and (2) simulation of the time evolution of the qubit state driven by the pulse sequence.    

In order to estimate the distortions to a microwave pulse introduced by two impedance mismatched elements we followed a standard approach of constructing an impulse response function $h(\texttau)$ for such a structure. Convolution of the input signal $x(t)$ with the structure’s impulse response function results in the output signal $y(t)$.

\begin{equation}
y\left ( t \right )=\int_{-\infty }^{\infty }h\left ( \tau  \right )x\left ( t - \tau  \right )d\tau
\end{equation}

As a first approach we constructed the impulse response function following the approach of Ref.~\citenum{saleh_valen}. 
The impulse response function is represented as a combination of delta functions separated by delay intervals, 

\begin{equation}
\label{eq:model1}
h\left ( t \right )= \sum_{k=0}^{\infty }B_{1}B_{2}\alpha ^{k}\left ( 1-\alpha  \right )\left ( \beta ^{k} -\beta ^{k+1}\right )\delta \left ( t-\tau_{k} \right ),
\end{equation} 
where the reflection coefficients of mismatched elements are defined as $\alpha$,$\beta = \pm 10^{-RL(\textbf{dB})/20}$ where $RL($dB$)$ is the return loss in decibels. The delay times are given by $\tau_{k} =L\left ( 2k+1 \right )v_{\textrm{p}}^{-1}$ and are determined by distance between the elements $L$ and the phase velocity $v_{\textrm{p}}$.  The amplitudes of the delta functions give the amplitudes of reflected signals. To check the model for consistency, we compare the results to a parallel approach that takes the inverse Fourier transform of a complex transmission function of a Fabry-P\'{e}rot-like structure \cite{ley_london},
 
\begin{equation}
\begin{split}
\label{eq:model2}
h\left ( t \right ) &= \frac{1}{2\pi }\int_{-\infty }^{\infty }S_{21}\left ( \omega  \right )d\omega \\
& =\frac{1}{2\pi }\int_{-\infty }^{\infty }\frac{d^{2}e^{-i\frac{\omega L}{c}}}{1-r^{2}e^{-2i\frac{\omega L}{c}}}e^{it\omega }d\omega,
\end{split}
\end{equation}
where $r$ and $d$ are reflection and transmission coefficients correspondingly. In practice, good agreement between the two approaches is achieved when Eq.~(\ref{eq:model1}) is summed to $k \sim 5$ reflections showing minor disagreement only in case of shorter pulses. The advantage of using the impulse response function Eq.~(\ref{eq:model1}) will be shown below.

Taking the above analysis further, we check how the distortion in the drive pulse affects gate fidelities. To accomplish this, we use a simple model of a resonantly driven spin qubit and numerically simulate its time evolution. In this model the state of the qubit evolves in time according to the following Hamiltonian:

\begin{equation}
\frac{H\left ( t \right )}{\hbar} = \frac{1}{2}\omega_{q} \left ( 1+\sigma _{z} \right )+\sigma _{x}A_{x}\cos \left ( \omega_{q} t+\phi  \right )
\end{equation} where $\omega_{q}$ is the qubit transition frequency, $A_{x}$ and $\phi$ are the amplitude and the phase of the drive signal, $\sigma _{x}$ and $\sigma _{z}$ are the Pauli matrices. In order to isolate the effects of drive pulse distortion on the qubit state time evolution the system is considered to be closed. 

The numerical simulation of qubit state time evolution is done using Qutip \--- an open-source python library for simulating the dynamics of quantum systems \cite{qutip_2}. Within the simulation as a form of bench-marking we apply all possible combinations of X and Y gates, also known as the ALLXY experiment in benchmarking of physical qubit systems \cite{gao-prxq-2021}. The end result is the fidelity between the state achieved for infinite return loss \textrho {}
and the state achieved with a given return loss \textsigma {}
defined as following:
\begin{equation}
F\left ( \rho ,\sigma \right )=\left | \left \langle \psi _{\rho }\mid \psi _{\sigma } \right \rangle \right |^{2}
\end{equation}
According to the simulation results based on impulse response function Eq.~(\ref{eq:model1}) the fidelity deviation $1-F$ depends periodically on the distance between the two unmatched elements (see Fig.~3 of the main article).  Since the simulation results for all combinations of X and Y gates were extremely close only the data for XY gate combination is presented on the plot. \ADDTXT{The length 0.276~meters chosen to show fidelity deviation dependence on return loss of the unmatched elements corresponds to a local maximum in the length dependence.} 

The periodical behavior of the fidelity deviation can be explained by analyzing how distortions in phase of the drive pulse affect the state trajectory on the Bloch sphere. A pulse passing through a transmission line with two mismatched elements will be a result of interference between the original pulse and the "ghosting" pulses that arise from partial reflections of the pulse between the mismatched elements. The resulting pulse may consist of sub-periods with different individual phase due to a delay in arrival of the component pulses that is defined by the distance between the two unmatched elements. The advantage of using the impulse response function Eq.~(\ref{eq:model1}) is that it allows to limit the amount of reflected pulses to a desired value and thus simplifying analysis of the mechanism affecting the qubit state. In a simplified case of a sub-period appearing as a result of interference of two overlapping pulses the resulting phase $\phi$ will depend on the phases of the interfering pulses $\theta _{1},\theta _{2}$ the following way:
\begin{equation}
\label{eq:phase}
\phi =\arctan \left ( \frac{A\sin\theta _{1}+B\sin\theta _{2}}{A\cos\theta _{1}+B\cos\theta _{2}} \right )
\end{equation}
  Depending on the proportion between the amplitudes $A,B$ of the interfering pulses (which would depend on the return loss of the mismatched elements) the resulting phase dependence changes form harmonic (amplitudes differ by an order of magnitude) to saw-tooth form (at close values of amplitudes). The change in phase and thus a change in state rotation vector-angle:
 \begin{equation} 
 \vec{\Omega} =\begin{pmatrix}
A_{x}\cos \phi\\ 
A_{x}\sin \phi\\ 
0
\end{pmatrix}
\end{equation}
 according to \ADDTXT{Eq.}~(\ref{eq:phase}) can be both positive and negative.
 Since changes to the rotation angle during state evolution lead to state trajectory curving, the resulting state point ends up at a distance from the intended one. And fidelity deviation $1-F$ between the resulting and intended state at small deviation values is equal to the square of the distance between the state points on the Bloch sphere. Thus periodic behavior of absolute value of the resulting phase in equation (\ref{eq:phase}) results in periodic behavior of the fidelity deviation. 

\bibliography{2portcal}
